# Temperature dependent As K-edge EXAFS studies of LaFe$_{1-x}$Co$_x$AsO (x = 0.0 and 0.11) single crystals

Boby Joseph[1,2], Alessandro Ricci[3], Nicola Poccia[1,4], Valentin G. Ivanov[5], Andrey A. Ivanov[5]. Alexey P. Menushenkov[5], Naurang L. Saini[6] and Antonio Bianconi[1,5*]

[1]*Rome International Centre for Material Science Superstripes, RICMASS, via dei Sabelli 119A, 00185 Rome, Italy*
[2]*Elettra, Sincrotrone Trieste, Strada Statale 14, Km 163.5, Basovizza, 34149 Trieste, Italy*
[3]*Deutsches Elektronen-Synchrotron (DESY), P10/PETRAIII - Coh. X-ray group, Notkestraße 85, 22607 Hamburg, Germany*
[4]*NEST Istituto Nanoscienze-CNR & Scuola Normale Superiore, Pisa, Italy*
[5]*National Research Nuclear University "MEPhI" (Moscow Engineering Physics Institute), Department of Physics of Solid State and Nanosystems, Kashirskoe sh. 31, 115409 Moscow, Russia*
[6]*Department of Physics, Sapienza University of Rome, P. le A. Moro 2, 00185 Rome, Italy*

*E-mail: antonio.bianconi@ricmass.eu

We report the experimental results of temperature dependent *polarized* As K-edge extended x-ray absorption fine structure (EXAFS) of LaFe$_{1-x}$Co$_x$AsO (x=0.0 and 0.11) single-crystals. By aligning the Fe-As bond direction in the direction of the x-ray beam polarization we have been able to identify an anomaly in the Fe-As bond correlations at the tetragonal to orthorhombic transition at 150K, while previous investigations with standard unpolarized EXAFS of undoped LaFeAsO powder samples were not able to detect any such anomaly. Using our approach we have been able to identify in the superconducting doped sample, LaFe$_{0.89}$Co$_{0.11}$AsO, a broad anomaly around 60 K. The low temperature anomaly has good correlations with the temperature dependence of several properties like resistivity, magnetic susceptibility, linear thermal expansion, *etc* indicating the emergence of the dynamical oscillations of the Fe - As pairs.

PACS numbers: 71.81.-g, 74.72.Dn ,64.60.-i,61.05.cj

## Introduction

The excitement in the condensed matter community on the iron-based superconductors (FeSC) continue to grow even after several years of intense worldwide research efforts, thanks to the several interesting experimental and theoretical investigations. The F-doped LaFeAsO systems [1] and the related lanthanide compounds [2] (so called 1111 systems) have attracted a renewed interest once quality single crystals of these systems become available [3-4], thus making way for experimental investigations [4-10]. These new experiments allow one to test the available theoretical predictions and also to give further experimental inputs needed for the possible formulation of an effective unified theory for the description of high temperature superconductors. The most important and puzzling characteristics of these compounds is the occurrences of magnetism and superconductivity in close vicinity in the phase diagram and the interplay between the two [10]. Magnetism in these systems have attracted a large theoretical interest [11-13], especially the unusually small magnetic moment associated with the Fe lattice. To obtain an overall agreement between the different theoretical tools and experimental observations, it was proposed that the 1111 systems posses a strong magnetic ground state, however with fluctuating domains which prelude the experimental detection Recently, using single crystals, in the NdFeAsO system, two additional phase transitions associated with the Nd and Fe magnetic moments were observed at low temperatures [14], indicating that indeed the rare-earth and Fe magnetic moments play more active role than those revealed by the earlier experiments [15-17]. Magnetic and superconducting transitions have a direct correlation with the changes in the electronic density of states, which in turn is driven by the subtle structural changes, in many cases directly visible as anomalies in the mean square relative displacements of the participating atomic bonds. Systematic local structural studies are necessary for the detection of such anomalies. In cuprates, such anomalies were detected in several systems using extended x-ray absorption fine structure (EXAFS) spectroscopy studies [18,19] and neutron diffraction based atomic pair distribution function (PDF) analysis [20,21]. EXAFS [22-29] and x-ray PDF [30-34] studies were also yielded several important information on the FeSC, however, lack of high quality single crystals limited those studies. Here we report a systematic EXAFS study on the LaFe$_{1-x}$Co$_x$AsO (x=0 and 0.11) system using millimetre-sized single crystals, which demonstrates the important lattice effects in determining the properties of the system.

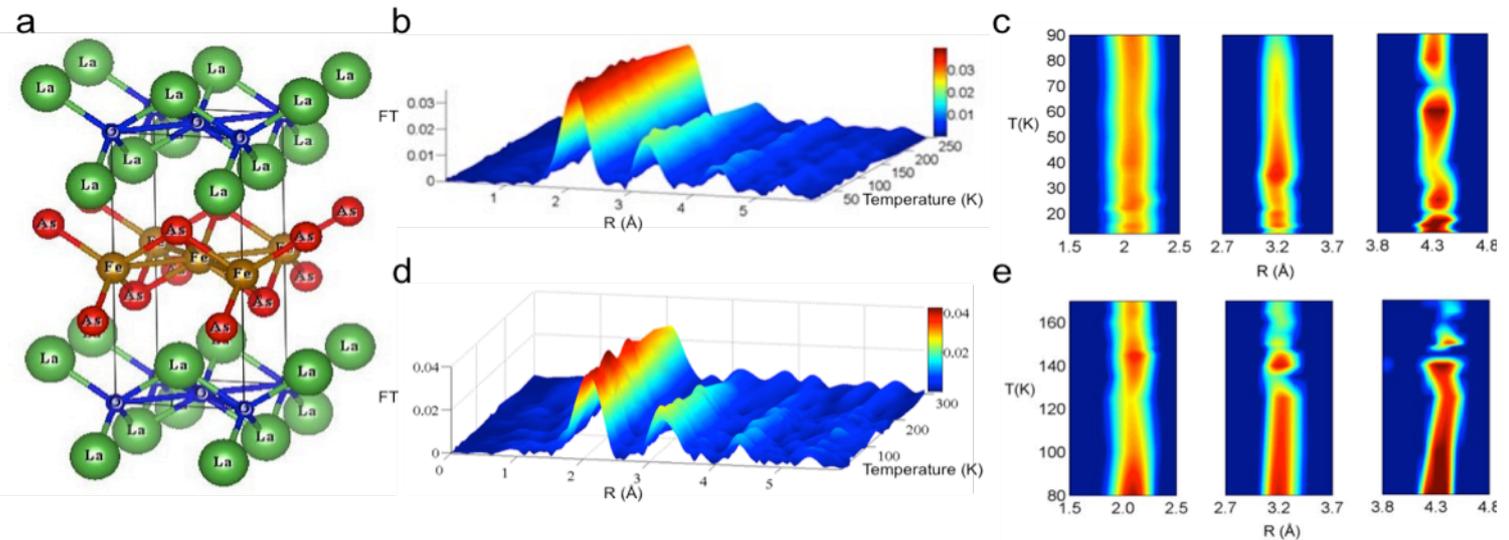

**Figure 1**: Unit cell of LaFeAsO (a). Fourier transform (FT) magnitudes of the As *K*-edge EXAFS oscillations at different temperatures for the LaFe$_{1-x}$Co$_x$AsO (x=0.11) (b, c) and LaFeAsO (d, e) single crystals. FTs are not corrected for the phase shifts, and represent raw experimental data.

**Materials and Methods**

LaFe$_{1-x}$Co$_x$AsO (x=0 and 0.11) single crystals of size around ~2 mm × 2 mm × 0.2 mm were grown under ambient pressure in NaAs flux [4]. Temperature dependent As $K$-edge EXAFS on these single crystals were carried out in fluorescence yield mode at the beamline, BM 29 of the European synchrotron radiation facility, Grenoble (France). Measurements were carried out between 15 and 300 K (with more than 25 temperature points for each crystals). The crystal was oriented so that the direction of the Fe-As bond is parallel to the electric field direction of the polarized x-ray photon beam. A continuous flow liquid He cryostat was used for the low temperature measurements. Sample temperature during measurements were monitored and controlled within ±1K. A minimum of 3 scans (many cases up to 5) were taken at each temperature on both samples. From the absorption spectra, EXAFS data were extracted following the standard procedures [35]. At each temperature, average of the different scans were used for the analysis.

EXAFS is a unique fast (with a measuring time scale of $10^{-15}$ s) local tool to get the instantaneous Fe-As bond length distribution without time averaging [18,19,35-39] which has first detected the short range charge density wave [18], the polaron size [37] and the nanoscale stripes phase [38] in cuprates. The structure of LaFeAsO is made of an alternate stacking of the [FeAs] active and spacer layers [LaO] [3,39] as shown in Fig. 1(a). The misfit strain [2,40] and the proximity of the chemical potential to an electronic topological Lifshitz transition [41,42] drive these layered system to an arrested nanoscale phase separation [43-44].

**Results and Discussion**

The raw experimental data (the Fourier transform magnitude) extracted from the As K-edge EXAFS in the complete temperature range is presented in Figs. 1(b) and (d). Instead a zoom over the three main peaks (representing the different atomic shells around the As) at two temperature ranges for the Co substituted and parent compounds are shown in Figs. 1 (c) and (e). Clear anomalies in the intensity are clear from this 2D colour plots in the Fourier transform peaks at 0.43 nm, close to the structural phase transition at 140 K in the parent compound, whereas the Co doped superconducting sample show clear anomalies around 60 K.

In case of the As $K$-edge EXAFS in La-1111 system, the first shell contribution involving As-Fe bonds is well separated from other contributions and thus a single shell modelling is very effective in extracting the quantitative Fe-As bond distributions [22,24,26]. The structure of LaFe$_{1-x}$Co$_x$AsO has a tetragonal symmetry at room temperature. For x=0.0, a structural transition to an orthorhombic phase appears below 150 K [2,6, 8-10]. For the As site

(probed by the As *K*-edge), there are four Fe near neighbours at a distance ~2.4 Å. The next nearest neighbours of As are La and O atoms. Contributions of these distant shells appear mixed, and it is very difficult to extract quantitatve information on those distant shells. However, the contribution of the Fe-As bonds are well separated from other contributions and can be analysed using a single shell fit to extract the quantiative information on bond correlations.

As mentioned earlier, the As *K*-edge EXAFS has been analyzed using a single shell fit to extract quatitative information on the Fe-As bond correlations. For the single shell fit, except the radial Fe-As distance and related mean square relative displacements (MSRD), describing the correlated Debye-Waller factor ($\sigma^2$), all other parameters (like the photo-electron energy origin, the number of near neighbors, *etc.*) were kept fixed in the conventional least squares modelling, using the phase and amplitude factors calculated using the FEFF program [26]. For such a two parameter fit, the number of independent data points for the present modelling: $N_{ind} \sim (2\Delta k\Delta R)/\pi$, where $\Delta k$ and $\Delta R$ are respectively the ranges in k and R space over which the data are analyzed [25], is 7 ($\Delta k=10$Å$^{-1}$ and $\Delta R=1.2$ Å). This makes the present analysis quite suitable for obtaining quantitative Fe-As bond correlations, minimising the correlation effects among the free parameters involved in the modelling.

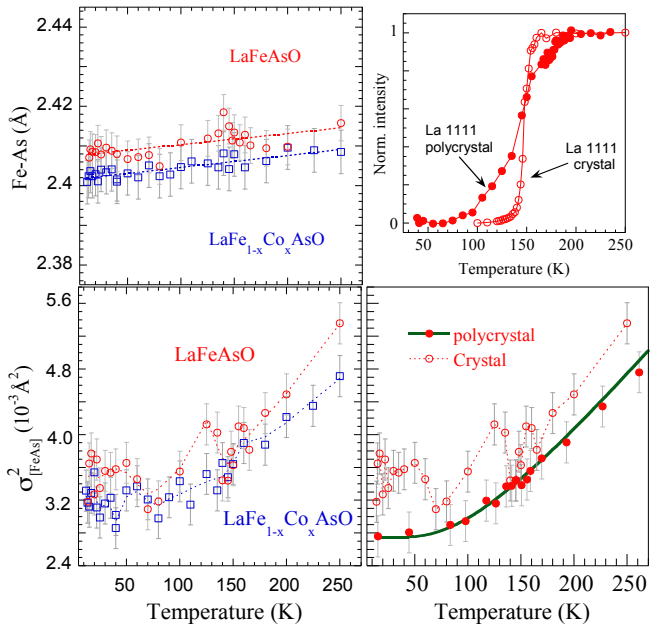

**Figure 2**: Variation of the Fe-As bond length and the corresponding mean square relative displacements (MSRD) with temperature for the LaFe$_{1-x}$Co$_x$AsO (x=0 and 0.11) single crystals extracted from the single shell modelling of the As K-edge EXAFS. Lower right panel shows the temperature dependence of the Fe-As MSRD for the LaFeAsO single crystal and polycrystalline powder [26] samples. Variation of the intensity of the diffraction reflection corresponding to the tetragonal (220) peak with temperature for the single crystal and polycrystalline powder samples of LaFeAsO [9] is shown in the right upper inset.

Figure 2, upper and lower left panels show respectively the temperature dependence of the Fe-As radial distance and the related mean square relative displacements (MSRD), describing the correlated Debye-Waller factor ($\sigma^2$), extracted from the above mentioned single shell modeling of the As *K*-edge EXAFS data. The Fe-As distance for the Co doped sample is slightly lower than the undoped sample in line with the reduced lattice parameters for the Co doped [47] sample.

Over-all the Fe-As bonds show weak temperature dependence (negligible linear thermal expansion). A careful look reveals the temperature dependence is not smooth, similar to the recent results from high resolution diffraction measurements [48]. The temperature dependence of the MSRD is different between the doped and undoped samples (figure 3, lower left panel). For the parent compound, the Fe-As MSRD show abrupt changes around 150 K. No such changes are seen in the Co doped sample. The orthorhombic distortion for our sample is reported to be occurring below 154.5 K, and the onset of magnetic order around 140 K. Both these transitions are absent in the Co doped sample, which shows a superconducting transition around 10 K.

Clearly, the anomalies seen in the Fe-As MSRD are related to the phase transitions in the system. Interestingly, such anomalies were not reported in the earlier EXAFS studies on these systems, where polycrystalline powder sample were used for the measurements. To highlight the difference, in figure 2 lower right panel, we present the temperature dependence of the Fe-As MSRD for the single crystal and the polycrystalline powder sample. The temperature dependence of the Fe-As MSRD is smooth for the polycrystalline power samples across the structural phase transition temperature [22,26], whereas for the single crystals, the Fe-As MSRD show clear anomaly. In a systematic diffraction study, Ricci *et al.*, highlighted the difference in the structural transition properties of the single crystals and the corresponding polycrystalline powders of the 1111 pnictides [9]. In Fig. 2, upper right panel, we show the variation of the intensity of the diffraction reflection corresponding to the tetragonal (220) peak with temperature for the single crystal and polycrystalline powder samples of LaFeAO [9]. In case of the polycrystalline powder sample, the structural transition is not sharp and extends over a temperature window of around 80 K [9]. Such a broad transition can mask the observation of the possible lattice anomalies of the participating bonds in the local structural measurements.

A further comparison of the temperature dependence of the Fe-As MSRD of the single crystal and polycrystalline powder (Fig. 2), shows that the MSRD of the single crystal also deviates from that of the polycrystalline powder below 60 K. In fact several properties of the LaFeAsO show a discontinuous behaviour both at the structural phase transition and at a lower temperature below 60 K. In Fig. 3, we make a comparison of the temperature dependence of the Fe-As MSRD with the resistivity [49], thermal expansion [50] and magnetic susceptibility [51] of the LaFeAsO. Very recent angle resolved photoemission spectroscopy (ARPES) measurements on the single crystal sample of LaFeAsO shows a gap opening after the structural phase transition [52]. The observed band shift (which happens around SPT and goes through SDW smoothly) is interpreted as spin density wave fluctuation (short range magnetic order) at the structural phase transition. Incidentally, from optical spectroscopy measurements, using single crystals [4], formation of a partial energy gap below the structural phase transition [52] and strong electron-phonon coupling were observed in the 1111 parent compounds [5]. The anomaly we observe around 150 K in the Fe-As bond correlations in LaFeAsO single crystal is mostly related to such a gap opening phenomena. It is worth recalling that the Cu-O bond length fluctuations observed in cuprate superconductors are identified as related to the pseudo-gap phenomena [18,19,21]. A recent ARPES study on CeFeAsO single crystal also showed the gap opening associated with the SPT/SDW transitions [8]. Temperature dependent pseudogap phenomena is observed in the LaFeAsO system by laser photo-emission studies [53]. The puzzling nature of magnetic and lattice phase transitions of FeSC is investigated via a first-principles Wannier function analysis of representative parent compound LaFeAsO. A rare ferro-orbital ordering is found to result in a highly anisotropic magnetic coupling, and drive both phase transitions—without resorting to widely employed frustration or nesting picture [54].

As is evident from figure 3, in the undoped sample, there is a clear anomaly in the MSRDs in the low temperature regime, in addition to the SPT. This low temperature anomaly clearly correlates with the observed low temperature discontinuity of resistivity (figure 3 upper panel), thermal expansion (figure 3 middle panel) and magnetic susceptibility (figure 3 lower panel). It is possible that fluctuating domain boundaries in a nanoscale phase separation scenario in iron based superconductors [55-58] like in cuprates [42-46] could explain the disagreement between theory and experiment in several crucial features of the FeSC.

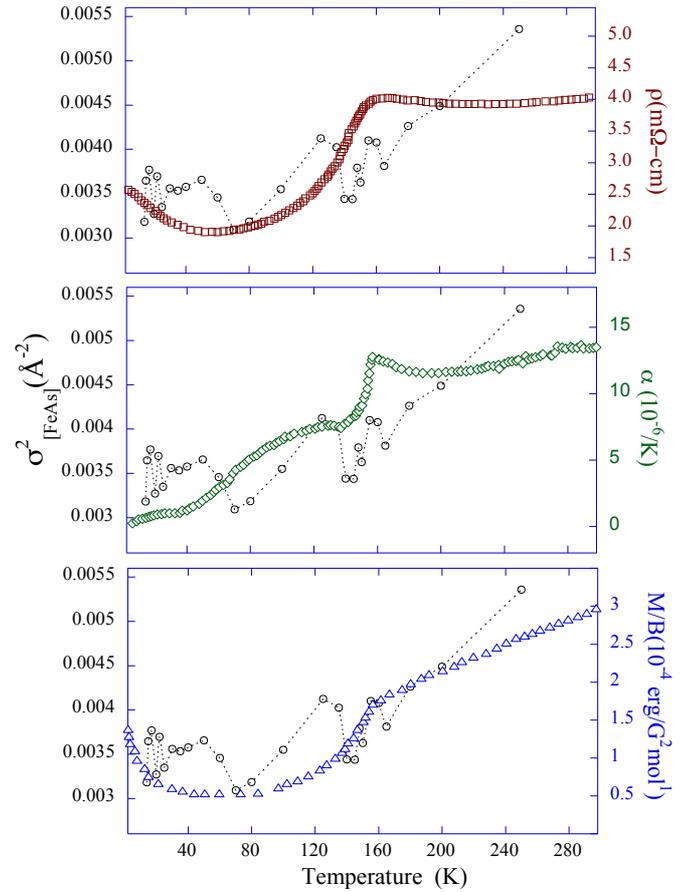

**Figure 3**: Variation of the Fe-As mean square relative displacements (MSRD) with temperature for the LaFeAsO single crystal extracted from the single shell modelling of the As K-edge EXAFS together with resisitvity [49], thermal expansion [50] and magnetization [51] data reported for the same system.

Indeed the magnetic ground state in the 1111 Fe systems have attracted a lot of theoretical interest [11-13,55], and it is seen that the moment ordering of the Fe in a ferro-magnetic state leads to destruction of magnetism. In view of the above studies, it is very tempting to attribute the second anomaly observed in the Fe-As MSRD to a possible low temperature charge density ordering in the parent compound. Interestingly, although weak in nature, (considering the noise level), such an anomaly seems to be also present in the doped sample as well (figure 3, lower left panel). In a recent susceptibility study, the interplay of magnetism and superconductivity in LaFeAsO$_{1-x}$F$_x$ is clearly shown [51]. These authors show that while antiferromagnetic SDW formation is suppressed by superconductivity, the data provide strong evidence for robust local antiferromagnetic correlations persisting even in the superconducting regime of the phase diagram. The charge distribution in RFeAsO$_{1-x}$F$_x$ (R=La, Sm) iron pnictides probed by arsenic nuclear quadrupole resonance indicate that the undoped and optimally doped or overdoped compounds feature a single charge environment, while two charge environments exists for the underdoped region [56]. From the temperature dependence of electron spin resonance in LaFeAsO$_{1-x}$F$_x$ ($x$=0 and 0.13), Wu *et al.,* suggest the existence of local moments in these materials [57]. The present results seem to support a nanoscale phase separation scenario in FeSC [25,41,42,51,56-60] which has been observed also in cuprates [42,43,44].

In conclusion, the temperature dependent Fe-As bond correlations of LaFe$_{1-x}$Co$_x$AsO (x=0 and 0.11) single-crystals studied using the As *K*-edge extended x-ray absorption fine structure (EXAFS) spectroscopy which show presence of anomalies in the Fe-As mean square relative displacements (MSRD) that were not seen in the corresponding polycrystalline powder sample. The anomalies of the FeAs MSRD around 150 K in LaFeAsO single-crystal is well correlated with the structural and spin density wave associated phase transitions. Such an anomaly is absent in the Co doped sample where no such long range structural or magnetic transitions exist. The absence any anomaly in Fe-As MSRD in the polycrystalline powder samples of LaFeAsO can be understood considering the difference in the sharpness of the phase transition in polycrystalline-powder and single-crystal [9]. Interestingly, both undoped and Co-doped LaFe$_{1-x}$Co$_x$AsO samples, show a low temperature anomaly around 60 K, with much weaker strength in the latter. The anomaly has a good correlation with the temperature dependence of the resistivity magnetic susceptibility, linear thermal expansion experiments indicating the importance of the structural effects in determining these properties. The results seems to support the phase separation scenario observed in several Fe based superconductors [25,41,42,51,56-60].

**Acknowledgements.** We thank Y-Q Yan, R. W. McCallum, and T. A. Lograsso for providing us the single crystals. We thank ESRF for beamtime allocation and Dr. O. Mathon of BM-23 for his help in the EXAFS data collection. V.G.I. and A.P.M. thank the Russian Science Foundation (project 14-22-00098) for support.

## References

1. Y. Kamihara, T. Watanabe, M. Hirano, and H. Hosono, *Journal of the American Chemical Society* 130, 3296 (2008), http://dx.doi.org/10.1021/ja800073m.
2. M. Fratini, R. Caivano, A. Puri, *et al.*, *Superconductor Science and Technology* 21, 092002+ (2008), http://dx.doi.org/10.1088/0953-2048/21/9/092002
3. J. Karpinski et al., *Physica C: Superconductivity* 469, 370 (2009).
4. J. Q. Yan *et al.*, *Applied Physics Letters* 95, 222504 (2009).
5. Z. G. Chen, R. H. Yuan, T. Dong, and N. L. Wang, *Physical Review B* 81, 100502 (2010), http://dx.doi.org/10.1103/PhysRevB.81.100502.
6. T. Dong, Z.G. Chen, R.H. Yuan, B. F. Hu, B. Cheng, N. L. Wang., *Physical Review B* 82, 054522 (2010). http://dx.doi.org/10.1103/PhysRevB.82.054522
7. A. Jesche, C. Krellner, M. de Souza, M. Lang, and C. Geibel, *Physical Review B* 81, 134525 (2010), http://dx.doi.org/10.1103/PhysRevB.81.134525.
8. H. Liu, G. F. Chen, W. Zhang, *et al.*, *Physical Review Letters* 105, 027001 (2010), http://dx.doi.org/10.1103/PhysRevLett.105.027001.
9. A. Ricci, N. Poccia, B. Joseph, *et al.*, *Physical Review B* 82, 144507 (2010), http://link.aps.org/doi/10.1103/PhysRevB.82.144507.
10. M. D. Lumsden and A. D. Christianson, *Journal of Physics: Condensed Matter* 22, 203203+ (2010), 8984, http://dx.doi.org/10.1088/0953-8984/22/20/203203.
11. T. Yildirim, *Physical Review Letters* 101, 057010+ (2008), ISSN 0031-9007, http://dx.doi.org/10.1103/PhysRevLett.101.057010.
12. Z. P. Yin, S. Lebègue, M. J. Han, B. P. Neal, S. Y. Savrasov, and W. E. Pickett, *Physical Review Letters* 101, 047001+ (2008), ISSN 0031-9007, http://dx.doi.org/10.1103/PhysRevLett.101.047001.
13. J. Wu, P. Phillips, and A. H. C. Neto, *Physical Review Letters* 101, 126401+ (2008), http://dx.doi.org/10.1103/PhysRevLett.101.126401.
14. W. Tian *et al.*, *Physical Review B* 82, 060514 (2010). http://dx.doi.org/10.1103/PhysRevB.82.060514
15. Y. Qiu, W. Bao, Q. Huang, *et al.*, *Physical Review Letters* 101, 257002+ (2008), http://dx.doi.org/10.1103/PhysRevLett.101.257002.
16. C. de la Cruz, Q. Huang, J. W. Lynn, *et al.*, *Nature* 453, 899 (2008), http://dx.doi.org/10.1038/nature07057.
17. J. Zhao *et al.*, *Nature Materials* 7, 953 (2008).
18. A. Lanzara, N. L. Saini, M. Brunelli, A. Valletta, and A. Bianconi, *Journal of Superconductivity and Novel Magnetism* 10, 319 (1997), http://dx.doi.org/10.1007/bf02765711.
19. N. L. Saini *et al.*, *Physical Review B* 55, 12759 (1997).
20. B. H. Toby, T. Egami, J. D. Jorgensen, and M. A. Subramanian, *Physical Review Letters* 64, 2414

(1990), http://dx.doi.org/10.1103/PhysRevLett.64.2414.
21. E. S. Božin, G. H. Kwei, H. Takagi, and S. J. L. Billinge, *Physical Review Letters* **84**, 5856 (2000), http://dx.doi.org/10.1103/PhysRevLett.84.5856.
22. C. J. Zhang, H. Oyanagi, Z. H. Sun, Y. Kamihara, and H. Hosono, *Physical Review B* **78**, 214513 (2008), http://dx.doi.org/10.1103/PhysRevB.78.214513.
23. A. Iadecola, S. Agrestini, M. Filippi, L. Simonelli, M. Fratini, B. Joseph, D. Mahajan, and N. L. Saini, *EPL (Europhysics Letters)* **87**, 26005 (2009), 0910.2329, http://dx.doi.org/10.1209/0295-5075/87/26005.
24. C. J. Zhang, H. Oyanagi, Z. H. Sun, Y. Kamihara, and H. Hosono, *Physical Review B* **81**, 094516 (2010), http://dx.doi.org/10.1103/PhysRevB.81.094516.
25. B. Joseph, A. Iadecola, A. Puri, L. Simonelli, Y. Mizuguchi, Y. Takano, and N. L. Saini, *Physical Review B* **82**, 020502 (2010), http://dx.doi.org/10.1103/PhysRevB.82.020502.
26. T. A. Tyson *et al.*, *Journal of Applied Physics* **108**, 123715 (2010). http://dx.doi.org/10.1063/1.3525999;
27. B.Joseph, A. Iadecola, L. Malavasi, N. L. Saini, *Journal of Physics: Condensed Matter* **23**, 112202 (2011) http://dx.doi.org/10.1088/0953-8984/23/26/265701;
28. B. Joseph, A. Iadecola, L. Simonelli, et al., *Superconductor Science and Technology* 26, 065005 (2013) http://dx.doi.org/10.1088/0953-2048/26/6/065005;
29. M. Y. Hascisalihoglu, E. Paris, B. Joseph et al., *Physical Chemistry Chemical Physics* 18, 9029 (2016), http://dx.doi.org/10.1039/C5CP07985C
30. D. Louca, K. Horigane, A. Llobet, R. Arita, S. Ji, N. Katayama, S. Konbu, K. Nakamura, T. Y. Koo, P. Tong, et al., *Physical Review B* **81**, 134524 (2010), http://dx.doi.org/10.1103/PhysRevB.81.134524.
31. L. Malavasi, G.A. Artioli, H. Kim, et al., *Journal of Physics: Condensed Matter* **23**, 272201 (2011) http://dx.doi.org/10.1088/0953-8984/23/27/272201;
32. B. Joseph, V. Zinth, M. Brunelli, B. Maroni, D. Johrendt, L. Malavasi, *Journal of Physics: Condensed Matter* **23**, 112202 (2011) http://dx.doi.org/10.1088/0953-8984/23/11/112202
33. J. L. Niedziela, M. A. McGuire, and T. Egami, *Physical Review B* 86, 174113(2012) http://dx.doi.org/10.1103/PhysRevB.86.174113
34. B. Joseph, C. Marini, N. Demitri, et al., *Superconductor Science and Technology* 28, 092001 (2015) http://dx.doi.org/10.1088/0953-2048/28/9/092001
35. R. Prins, D. Koningsberger, Ed., *X-ray Absorption: Principles, Applications, Techniques of EXAFS, SEXAFS, XANES*, (Wiley, New York, 1988).
36. A. Bianconi and R. Bachrach, *Physical Review Letters* 42, 104 (1979), http://dx.doi.org/10.1103/physrevlett.42.104.
37. A. Bianconi, M. Missori, H. Oyanagi, H. Yamaguchi, Y. Nishiara, and S. Della Longa, *Europhysics Letters (EPL)* 31, 411 (1995), http://iopscience.iop.org/0295-5075/31/7/012.
38. A. Bianconi, N. L. Saini, A. Lanzara, M. Missori, T. Rossetti, H. Oyanagi, H. Yamaguchi, K. Oka, and T. Ito, *Physical Review Letters* **76**, 3412 (1996), http://dx.doi.org/10.1103/PhysRevLett.76.3412.
39. A. Ricci, B. Joseph, N. Poccia, W. Xu, D. Chen, W. S. Chu, Z. Y. Wu, A. Marcelli, N. L. Saini, and A. Bianconi, *Superconductor Science and Technology* **23**, 052003+ (2010), http://dx.doi.org/10.1088/0953-2048/23/5/052003
40. N. Poccia, A. Ricci, and A. Bianconi, *Advances in Condensed Matter Physics* **2010**, 261849 (2010), http://dx.doi.org/10.1155/2010/261849
41. R. Caivano, M. Fratini, N. Poccia, et al., *Superconductor Science and Technology* 22, 014004 (2009), http://dx.doi.org/10.1088/0953-2048/22/1/014004
42. A. Bianconi, *Nature Physics* 9, 536 (2013), ISSN 1745-2473, http://dx.doi.org/10.1038/nphys2738.
43. A. Bianconi, *International Journal of Modern Physics B* 14, 3289 (2000), http://dx.doi.org/10.1142/S0217979200003769
44. G. Campi, A. Bianconi, N. Poccia, et al., *Nature* 525, 359 (2015), http://dx.doi.org/10.1038/nature14987
45. K. Kugel, A. Rakhmanov, A. Sboychakov, N. Poccia, and A. Bianconi, *Physical Review B* 78, 165124 (2008), http://dx.doi.org/10.1103/physrevb.78.165124
46. A. Bianconi, N. Poccia, A. O. Sboychakov, A. L. Rakhmanov, and K. I. Kugel, *Superconductor Science and Technology* 28, 024005+ (2015), http://dx.doi.org/10.1088/0953-2048/28/2/024005
47. C. Wang *et al.*, *Physical Review B* **79**, 054521 (2009), http://dx.doi.org/10.1103/PhysRevB.79.054521
48. N. Qureshi, Y. Drees, J. Werner, S. Wurmehl, C. Hess, R. Klingeler, B. Büchner, M. T. F. D'iaz, and M. Braden, *Physical Review B* **82**, 184521+ (2010), http://dx.doi.org/10.1103/PhysRevB.82.184521.
49. M. A. McGuire *et al.*, *Physical Review B* **78**, 094517 (2008).
50. L. Wang *et al.*, *Physical Review B* **80**, 094512 (2009), http://dx.doi.org/10.1103/PhysRevB.80.094512.

51. R. Klingeler, N. Leps, I. Hellmann, A. Popa, U. Stockert, C. Hess, V. Kataev, H. J. Grafe, F. Hammerath, G. Lang, et al., *Physical Review B* **81**, 024506 (2010), http://dx.doi.org/10.1103/PhysRevB.81.024506.
52. C. Liu *et al.*, *Physical Review B* **82**, 075135 (2010). http://dx.doi.org/10.1103/PhysRevB.82.075135
53. Y. Ishida, T. Shimojima, K. Ishizaka, *et al.*, *Physical Review B* **79**, 060503+ (2009), http://dx.doi.org/10.1103/PhysRevB.79.060503.
54. C. C. Lee, W. G. Yin, and W. Ku, *Physical Review Letters* **103**, 267001 (2009), http://dx.doi.org/10.1103/PhysRevLett.103.267001.
55. M. J. Han, Q. Yin, W. E. Pickett, and S. Y. Savrasov, *Phys. Rev. Lett.* **102**, 107003 (2009), http://dx.doi.org/10.1103/PhysRevLett.102.107003.
56. G. Lang, H. J. Grafe, D. Paar, *et al.*, *Physical Review Letters* **104**, 097001 (2010), http://dx.doi.org/10.1103/PhysRevLett.104.097001.
57. T. Wu, J. J. Ying, G. Wu, R. H. Liu, Y. He, H. Chen, X. F. Wang, Y. L. Xie, Y. J. Yan, and X. H. Chen, *Physical Review B* 79, 115121 (2009), http://dx.doi.org/10.1103/PhysRevB.79.115121
58. J. T. Park, D. S. Inosov, C. Niedermayer, *et al.*, *Physical Review Letters* **102**, 117006 (2009), http://dx.doi.org/10.1103/PhysRevLett.102.117006;
59. M. Bendele, A. Barinov, B. Joseph *et al*., *Scientific Reports* 4. 5592 (2014) http://dx.doi.org/10.1038/srep05592
60. A. Ricci, N. Poccia, B. Joseph et al., *Physical Review B* 91, 020503 (2015) http://dx.doi.org/10.1103/PhysRevB.91.020503